\documentclass[aps,prl,superscriptaddress,reprint,showpacs,noeprint]{revtex4-1} 

\usepackage[utf8]{inputenc}

\usepackage{subfigure}

\usepackage{graphicx}

\usepackage{verbatim}

\usepackage{amsmath}
\usepackage{amssymb}
\usepackage{bm}
\usepackage{braket}

\begin{document}

\title{Super-Resolution Quantum Imaging at the Heisenberg Limit}
\author{Manuel Unternährer}
\email{manuel.unternaehrer@iap.unibe.ch}
\author{Bänz Bessire}
\affiliation{Institute of Applied Physics, University of Bern}
\author{Leonardo Gasparini}
\author{Matteo Perenzoni}
\affiliation{Fondazione Bruno Kessler FBK, Trento}
\author{André Stefanov}
\affiliation{Institute of Applied Physics, University of Bern}

\date{\today} 

\begin{abstract}
Quantum imaging exploits the spatial correlations between photons to image object features with a higher resolution than a corresponding classical light source could achieve. Using a quantum correlated $N$-photon state, the method of optical centroid measurement (OCM) was shown to exhibit a resolution enhancement by improving the classical Rayleigh limit by a factor of $1/N$. In this work, the theory of OCM is formulated within the framework of an imaging formalism and is implemented in an exemplary experiment by means of a conventional entangled photon pair source. The expected resolution enhancement of a factor of two is demonstrated. The here presented experiment allows for single-shot operation without scanning or iteration to reproduce the object in the image plane. Thereby, photon detection is performed with a newly developed integrated time-resolving detector array. Multi-photon interference effects responsible for the observed resolution enhancement are discussed and possible alternative implementation possibilities for higher photon number are proposed.
\end{abstract}

\pacs{42.50.-p, 42.50.St, 42.50.Dv, 42.50.Xa}

\maketitle 


\section{Introduction}

In metrology, the optimal measurements of a parameter under restricted use of limited measurement resources are studied \cite{Kapale07,Gio11,Simon17}. 
Using $N$ independent particles for probing a sample,  the parameter estimation error is improved by $1/\sqrt{N}$ beyond what can be achieved using a single particle, and is called standard quantum limit (SQL)  \cite{Gio06}. 
It was shown that the best possible measurement strategy with $N$ particles  is by using  quantum correlated states. This leads to an  $1/N$ improvement in estimation error \cite{Gio06}, an optimum which is called Heisenberg limit  \cite{Ou97}. In the case of an interferometric parameter estimation using photons, the multi-photon states at wavelength $\lambda$ exhibit features described by the de~Broglie wavelength $\lambda/N$ 
\cite{Jac95}. 
In imaging, i.e. the transmission of object shape information to an image plane,  the wavelength $\lambda$ of the used illumination and the numerical aperture (NA) of the imaging system determine image resolution through the Rayleigh resolution limit $1.22\lambda/\text{NA}$ \cite{Good88}. By changing the illumination to a spatially correlated or even entangled light source, only improvements at the SQL are achieved  \cite{Gio09, Santos08, Guerr10, Xu15}. 
Reaching the Heisenberg limit using temporally or spectrally shaped classical light and non-linear multi-photon absorbers systems for detection is shown in 
\cite{Yablo99,Peer04, Hemmer06} for interference fringes. A scheme using monochromatic quantum light for super-resolution lithography was proposed in 
\cite{Boto00, Bjork01} which iteratively builds up in principle arbitrary structures.

Despite these advances, no actual imaging of object features at the Heisenberg limit has been performed yet. One reason is the lack of  $N$-photon transmitters. These are shown in \cite{Gio09} 
to allow for imaging at the Heisenberg limit but can be omitted by preparing a quantum state which would be fully transmitted by such a device, as realized in this work. Furthermore,  time-consuming scanning or iteration in generation of the optical states  to build up image structures in the aforementioned  schemes prevented their application. Our experimental implementation of the state generation operates in a single-shot mode. Moreover, quantum imaging was hindered by the low speed of correlations measurements using scanning single-pixel devices. A recently developed integrated sub-nanosecond time-resolving \mbox{2-D} detector array allows for fast correlation measurements without scanning \cite{Gasp18}.
Most notably, detection efficiency is highly enhanced by OCM.

Tsang presented in \cite{Tsang09} the OCM method  and showed its ability to efficiently transmit the centroid position of a monochromatic, spatially entangled $N$-photon state beyond the Rayleigh diffraction limit. He theoretically showed that the resolution enhancement scales with $1/N$ corresponding to the Heisenberg limit where the de~Broglie wavelength $\lambda/N$ of the photons  determines the resolution  and not the wavelength $\lambda$ of the individual photons appearing in the classical Rayleigh limit. OCM was first implemented experimentally in \cite{Shin11} for $N=2$ and compared to before known quantum lithographic techniques. Furthermore, it was implemented in \cite{Rozema14} 
for photon numbers $N=2$~to~$4$. Both groups demonstrate super-resolution at the Heisenberg limit by measuring oscillation periods of interference fringes of two plane waves.

In this work, the OCM method is used in an imaging setting where  actual object features instead of interference fringes are observed. The OCM result is derived  for a general imaging system. Moreover, coherent OCM imaging  is experimentally implemented for photon number $N=2$, where an experimental setup is presented which allows to generate an entangled biphoton state containing the super-resolved OCM image.


\section{Theory}

An object can be described by its transmission aperture function $A(\bm\rho)$ in transverse position coordinates $\bm\rho = (x,y)$. 
For a monochromatic, spatially coherent and uniform light source at wavelength $\lambda$, the electric field after the object becomes  $E(\bm\rho) \propto A(\bm\rho)$. The goal of an imaging system is to reproduce the object field distribution in a distant plane where it can be measured or exposes a film.
For a general imaging system, the field intensity in the image plane reads
\begin{equation}\label{Eq:ImgCoh}
I(\bm\rho) = \left| \int d^2\!\bm\rho' \; A(\bm\rho') \, h\left(\frac{\bm\rho}{m} - \bm\rho'\right) \right|^2 = \left|\left(A \ast h\right)\left(\frac{\bm\rho}{m}\right)\right|^2
\end{equation}
with  magnification $m$ and  point-spread function (PSF) $h(\bm\rho)$ specific to that system and used wavelength \cite{Shih07,Good88}. A translation invariant PSF is assumed. Image resolution is then given by the width of the PSF and  an optimal image is achieved with $h(\bm\rho)= \delta^{(2)}(\bm\rho)$.

In order to transmit object features with an imaging system below the size of its PSF, i.e. the Rayleigh limit, Tsang proposed to replace the classical field distribution at the object plane by quantum correlated multi-photon states \cite{Tsang09}.
From these results, an explicit state can be constructed for the case of a coherent image of the object $A(\bm\rho)$. 
For $N$ photons at wavelength $\lambda$, the OCM image state reads
\begin{equation} \label{Eq:PsiNF}
\ket{\Psi} = \int d^2\!\bm\rho_1 \ldots d^2\!\bm\rho_N \; A\left(\frac{\bm\rho_1+\ldots+\bm\rho_N}{N}\right)\ket{\bm\rho_1, \ldots,  \bm\rho_N}
\end{equation}
in transverse positions of the photons $\bm\rho_{1},\ldots,\bm\rho_N$. 
Introducing new coordinates simplifies further analysis. The centroid position $\bm X$ and the deviations $\bm\xi_k$ are defined by
$$
	\bm X = \frac 1 N \sum_{k=1}^N \bm\rho_k, \quad \bm\xi_k = \bm\rho_k - \bm X, \; k\in\{1,\ldots, N\}.
$$
Due to  $\sum_k \bm\xi_k = 0$, the $N$-tuple $(\bm X,\bm\xi_1, \ldots, \bm\xi_{N-1})$ forms a complete coordinate system. In these coordinates, the quantum state encodes the image in the centroid position $\bm X$. Notice the infinite extension of the state in all $\bm\xi_k$ coordinates in this optimal case.

Propagating the electric field from the object plane trough the imaging system, the $N$-photon detection probability density in the image plane is given by the $N$-th order correlation function~\cite{Abb02}
\begin{multline*}
G^{(N)}(\bm\rho_1,\ldots,\bm\rho_N) = \left| \int d^2\!\bm\rho_1' \ldots d^2\!\bm\rho_N' \; A\left(\bm X' \right)  \right. \\
\left. h\left(\frac{\bm\rho_1}{m}-\bm\rho_1'\right) \ldots h\left(\frac{\bm\rho_N}{m}-\bm\rho_N'\right)  \right|^2.
\end{multline*}
A change to the coordinates $\bm X $ and $\bm\xi_k$ with  $k \in \{1,\ldots , N-1\}$ leads to
\begin{equation}\label{Eq:OCMG2}
G^{(N)}(\bm X, \bm\xi_1, \ldots, \bm\xi_{N-1}) =  \left|(A \ast H)\left(\frac{\bm X}{m}\right)\right|^2,
\end{equation}
where  the centroid PSF given by the $N$-times repeated self-convolution
\begin{equation}\label{Eq:Hrho}
H(\bm X) =   N^2 \,(\underbrace{h \ast  \ldots \ast h}_{\times N})(N\bm X) 
\end{equation}
determines the resolution of the image in the centroid coordinate $\bm X$. As formally explicit in comparison to Eq.~\eqref{Eq:ImgCoh}, the image is formed coherently. By replacing Eq.~\eqref{Eq:PsiNF} with an appropriate mixed state, an incoherent imaging variant can be derived as well, see supplemental material~\cite{SuppMat}. Summing $G^{(N)}$ over the $\bm\xi_k$ coordinates in dependence of the $\bm X$ coordinate yields the  \mbox{2-D} image. Due to this property, every $N$-fold coincident event carries image information and results thereby in an increased efficiency compared to methods where only spatially coincident photons are used. 
 
In the following, we consider for the imaging system a single lens with a circular pupil of radius $R$ determining its NA and Rayleigh resolution limit.
The PSF is then given by a sombrero function  $h(\bm\rho) = \operatorname{somb}(2\pi R|\bm\rho|/s_o\lambda)$ with the distance $s_o$ from the  object plane to the lens \cite{Shih07,Good88}.
By assuming $s_o \gg |\bm\rho|^2/\lambda$ for $\bm \rho$ in the object and detector area, the phase  $\exp(i\pi|\bm\rho|^2/\lambda s_o)$ appearing in the object plane for single lens imaging can be neglected. This establishes the  translation invariance of the PSF.
Using Eq.~\eqref{Eq:Hrho}, it can be shown that  
\begin{equation}\label{Eq:SingleLensH}
H(\bm X) = C \operatorname{somb}\left(\frac{2\pi R N}{s_o\lambda}|\bm X| \right)
\end{equation}
with an appropriate normalization constant $C$.
The additional factor of $N$ in the argument  reduces the width of the PSF, a spatial resolution enhancement corresponding to a Heisenberg $1/N$ scaling in photon number $N$.
 This is the result of \cite{Tsang09} expressed in an imaging formalism.
Eq.~\eqref{Eq:SingleLensH} can be understood in terms of a Fourier transform version of Eq.~\eqref{Eq:Hrho}. With the Fourier transform $\tilde h(\bm q)$ of $h(\bm\rho)$  in transverse wavevector coordinates $\bm q$,  Eq.~\eqref{Eq:Hrho} is equivalent to
$$
H(\bm X) =  \frac{N^2}{(2\pi)^2} \int d^2\!\bm q \; \left(\tilde{h}(\bm q)\right)^{N} \, e^{i N \bm q \cdot \bm X}. 
$$
As $\tilde h(\bm q)$ is given by the lens pupil function \cite{Good88}, any power of it is of unity transmission amplitude within its circular region. This results in a  $N$-times narrower but otherwise equal PSF as for classical imaging. Note that for a Gaussian  pupil function (apodization), the centroid PSF narrows only with $1/\sqrt{N}$, corresponding to the SQL.

For discussing the mechanism of resolution enhancement, it is of value to determine the photon correlations in the plane of the lens pupil. Assuming  $s_o$ to be large,  correlations in far-field can be considered. The OCM state of Eq.~\eqref{Eq:PsiNF}  in a far-field basis is given by
\begin{equation}\label{Eq:PsiFF}
\ket{\Psi} = \int d^2 \! \bm q  \; \tilde A \left(N{\bm q}\right)\ket{\bm q,\ldots, \bm q}
\end{equation}
where $\tilde A(\bm q)$ is the Fourier transform of $A(\bm\rho)$ in the transverse wavevector coordinate $\bm q$. This coordinate can be related to a position in the pupil plane by $\bm\rho = ( s_o \lambda /2\pi) \bm q $ in paraxial approximation \cite{Good88}. Strong position correlation   is therefore present at the pupil. The prefactor in the argument produces a $N$-times narrower far-field diffraction pattern as it is observed for classical illumination at the de~Broglie wavelength. 


\section{Experiment}

\begin{figure}[tp]
\centering
\includegraphics[scale=0.8]{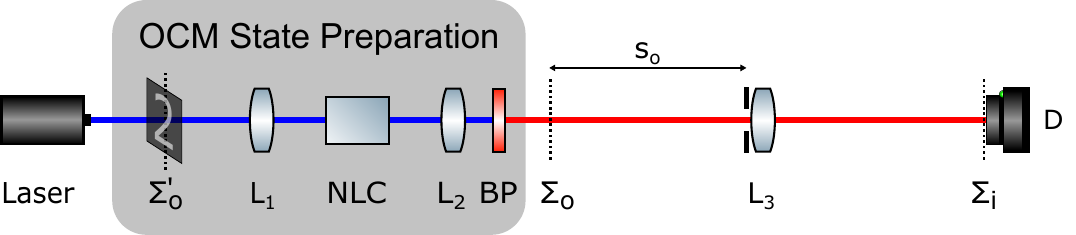}
\caption{(color online) The optical setup consisting of OCM state preparation and single-lens imaging (right). A 405\,nm laser source illuminates an object in the plane $\Sigma'_o$. 4-f imaging from  $\Sigma'_o$ to the output plane $\Sigma_o$ is performed by the lenses $L_1$ and $L_2$. The non-linear crystal (NLC) in the central far-field plane produces photon pairs in SPDC. A bandpass filter (BP) transmits 810\,nm. The actual imaging under investigation is performed by the lens $L_3$ from output plane $\Sigma_o$ to image plane $\Sigma_i$. $L_3$ is equipped with a circular pupil to reduce its NA and thereby the resolution of the imaging system. The 2-D detector array $D$ measures spatial biphoton correlations. }
\label{fig:Setup}
\end{figure}

Our exemplary experimental implementation of OCM imaging requires first the preparation of the quantum state of Eq.~\eqref{Eq:PsiNF} which is then followed by the actual imaging under investigation.
The optical setup for the OCM state preparation with photon number $N=2$ is depicted in Fig.~\ref{fig:Setup}. An object aperture 
$A(\bm\rho)$ is illuminated by a CW pump laser at 405\,nm and 30\,mW in the plane $\Sigma'_o$. A \mbox{4-f} lens system, consisting of two lenses $L_1$ and $L_2$ of focal length $f=50$\,mm, images the object to the preparation output plane $\Sigma_o$. In the far-field plane  between the lenses, a 5\,mm long periodically poled  KTiOPO$_4$  non-linear crystal (NLC) generates photon pairs at  $810$\,nm in type-0, frequency-degenerate, collinear spontaneous parametric down-conversion (SPDC). 

Using the approximation of a thin crystal and a plane wave pump, the entangled biphoton state in the output plane $\Sigma_o$ is derived in 
\cite{Abb02} and is given by Eq.~\eqref{Eq:PsiNF} for $N=2$. In the thick crystal case, it can be shown to read
\begin{equation}\label{Eq:PsiSinc}
	\ket{\Psi} = \int d^2\!\bm\rho_1 d^2\!\bm\rho_2 \, A\left(-\frac{\bm\rho_1+\bm\rho_2}{2}\right)\,
	\operatorname{sinc}\left(\frac{  \Delta k \,L}{2} \right) \ket{\bm\rho_1,\bm\rho_2}
\end{equation}
with the NLC length $L$ and  wavevector mismatch $\Delta k(\bm q_1, \bm q_2)$ \cite{Monken98} evaluated at $\bm q_k = (2\pi/\lambda f)\bm\rho_k,\;k \in \{ 1,2\}$. The mentioned infinite extension of the ideal OCM state in Eq.~\eqref{Eq:PsiNF} is here restricted by the dependence of this phase-matching function on $\bm\xi_1 = (\bm\rho_1-\bm\rho_2)/2$. For the experimental parameters, a FWHM of $1.1$\,mm can be calculated along this coordinate. This finite value is much larger then the width of the imaging PSF and will thereby not limit image resolution.

The imaging under investigation from $\Sigma_o$ to the image plane $\Sigma_i$ is performed with the lens $L_3$ of $250$\,mm focal length  at a distance $s_o=355$\,mm between $\Sigma_o$ and the lens. With the given detector and object sizes, the assumption of a large $s_o$ in the theoretical section is valid. A magnification of $m=2.4$ is measured. A circular pupil of radius  $R$ in front of the lens limits the NA and thereby the resolution of the image. For the used low NA, where the wavelength dependence on resolution is measurable with the given detector pixel size, an aperture $R = 1.38$\,mm is chosen. 

The OCM state is spectrally filtered by a bandpass filter (BP)  transmitting $810\pm5$\,nm. Its imaging capability is compared to classical light sources. These are spatially coherent, monochromatic illumination at 405\,nm and 810\,nm, as well as spatially incoherent light at 810\,nm. The former are implemented using collimated lasers, the latter by a halogen incandescent lamp spectrally filtered at $810\pm5$\,nm. For imaging with classical light, the setup of Fig.~\ref{fig:Setup} is used unaltered and allows comparison with otherwise identical parameters. Only the light source and spectral filters have to be replaced.

The newly developed detector used in the image plane $\Sigma_i$ is an integrated, fully digital 32$\times$32 pixel sensor array with single-photon sensitivity  manufactured in CMOS technology \cite{Gasp18}. This device contains for every pixel a dedicated time-to-digital converter which timestamps the first detection event at 205\,ps resolution in a  frame of 45\,ns duration. At an observation rate  of $800$\,kHz, a measurement duty-cycle of   $3.6\%$ is achieved. The photon detection efficiency reaches 5\% at 400\,nm and 0.8\% at 810\,nm. 
 The median dark count rate is below 1\,kHz over the pixel population. 
 Covering a sensitive region of $1.4\times 1.4$\,mm$^2$, the sensor is capable of  efficiently measuring second-order correlation functions similarly to its  predecessor presented in \cite{Unt16}. Despite these type of devices exist since 2009 \cite{Gasp18}, their use in quantum optics applications were limited by low fill-factor (1-5\%). Only recently it was possible to achieve  19.48\%   allowing useful detection efficiencies \cite{Gasp18}. For this work, a coincidence window of 1\,ns is used where accidental events including all dark counts can be measured and removed. 

\medskip

\begin{figure}
\subfigure[OCM 810\,nm]{
   \includegraphics[height=4cm]{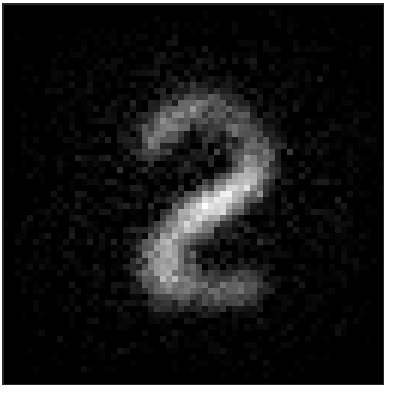}\label{fig:Digit2D3}}\hspace{0cm}
\subfigure[810\,nm]{
   \includegraphics[height=4cm]{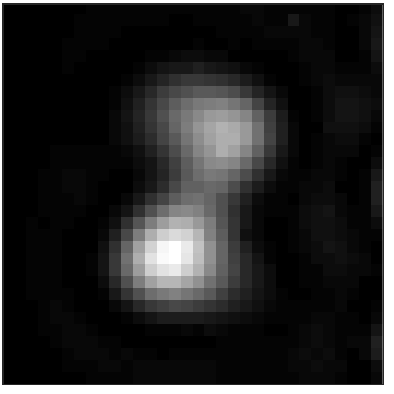}\label{fig:Digit2_810}}\hspace{0cm}\\
\subfigure[405\,nm]{
   \includegraphics[height=4cm]{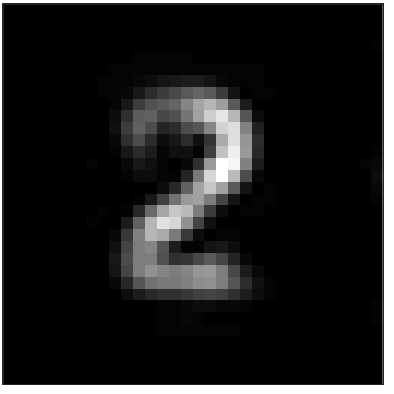}\label{fig:Digit2_405}}\hspace{0cm}
\subfigure[810\,nm, incoherent]{
   \includegraphics[height=4cm]{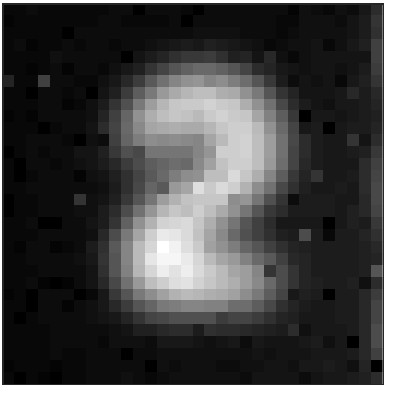}\label{fig:Digit2_810inc}}\hspace{0cm}
\caption{Imaging of an object using a single lens with different illumination light sources.  The same low NA  is used to demonstrate the wavelength dependence of resolution. Spatially coherent laser illumination is used in (b) and (c).  The region of $1.4\times1.4$\,mm$^2$ is acquired by a $32\times32$ pixels sensor. Biphoton OCM yields images at half-pixels and achieves a comparable image resolution at 810\,nm as a coherent light at 405\,nm.}
\label{fig:Digit2}
\end{figure}

In Fig.~\ref{fig:Digit2}(a--d), a $200\times300\,\mu$m object aperture is imaged at low NA values with different light sources. In the case of OCM, the measurement of the full second-order correlation function of Eq.~\eqref{Eq:OCMG2} yields the image of Fig.~\ref{fig:Digit2D3} by summing over $\bm\xi_1$. This introduces a vignetting effect in the resulting image due to the finite sensor size which acquires a broader range of $\bm\xi_1$ values in the image center than at the edges. It can be avoided by averaging  instead of summing over the available $\bm\xi_1$ values but would lead to higher statistical noise at the edges. 
Because the image is encoded in the centroid position, the image can be reconstructed at half-pixel precision leading to $63\times 63$ pixels images acquired by the 32$\times$32 pixels sensor.  In order to  suppress detector  crosstalk between adjacent pixels, only events with $|\bm\xi_1|> 1$\,pixel are considered. A measurement time of ten hours are used at low NA in Fig.~\ref{fig:Digit2D3}. 

\begin{figure}
\centering
\includegraphics[scale=.75]{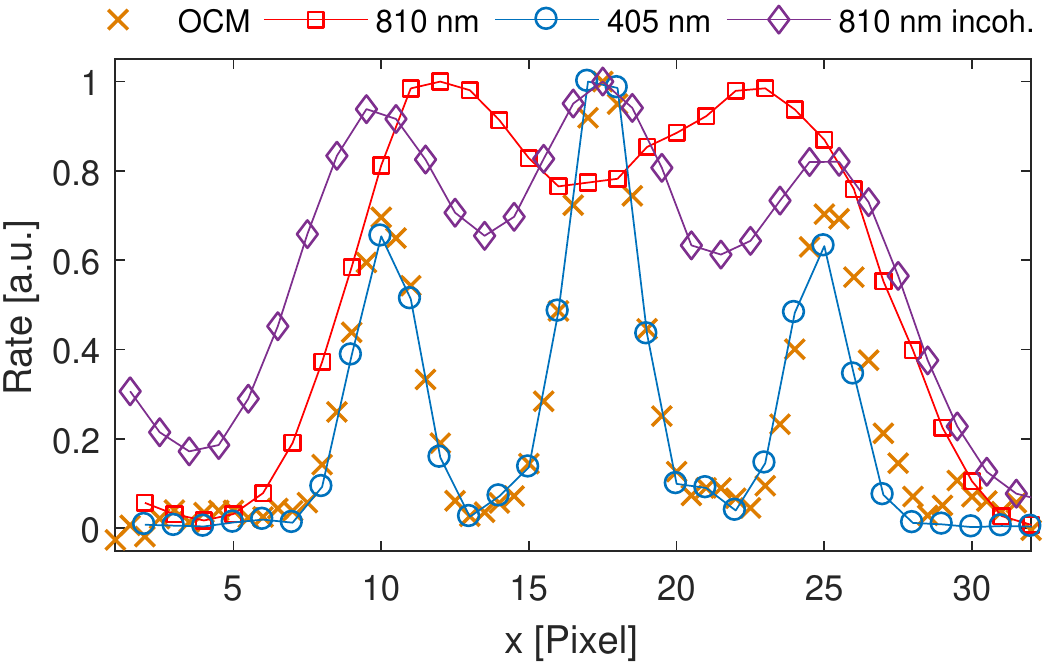}
\caption{(color online) A triple-slit object of 70\,$\mu$m line width is used for resolution comparison. A crossections of the image at low NA using biphoton OCM at 810\,nm (crosses), spatially coherent illumination at 810\,nm (squares) and 405\,nm (circles), and incoherent light at 810\,nm (diamonds). OCM shows an advantage which is  practically identical to the double resolution given with 405\,nm. }
\label{fig:TripleSlit}
\end{figure}

\begin{figure}
\centering
\includegraphics[scale=.75]{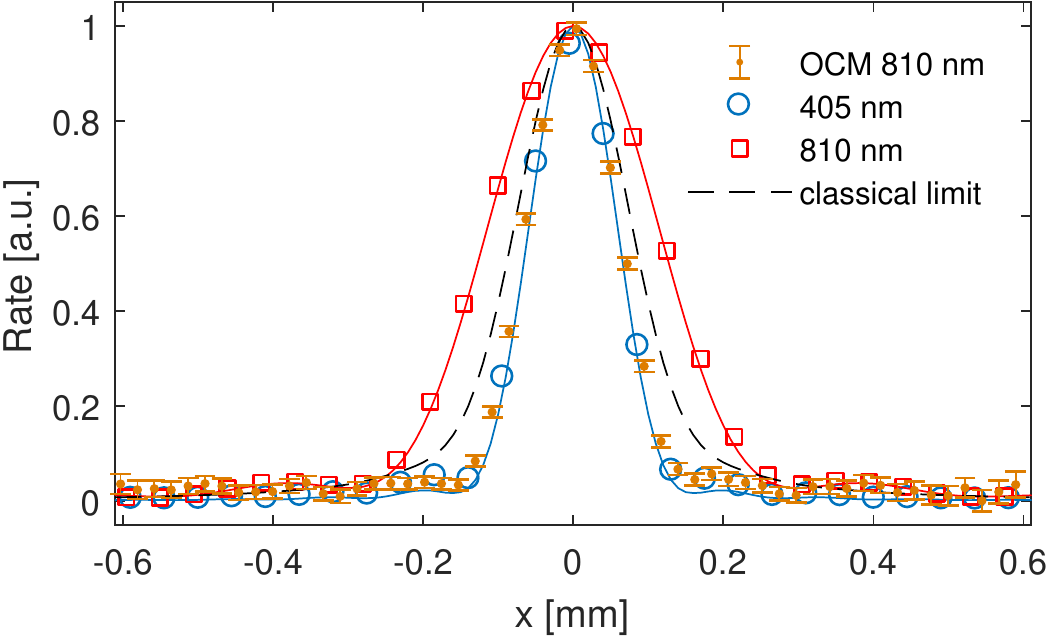}
\caption{(color online) Projection of the PSF of low NA single lens imaging  at different light sources. Measurements with coherent light at 810\,nm (squares) and 405\,nm (circles) are shown with their theoretical curves. The biphoton OCM PSF at 810\,nm (dots) closely agrees with 405\,nm and confirms its doubled resolution. Statistical $2\sigma$ errors are shown. The OCM PSF surpasses a theoretical, SQL-scaling PSF of the centroid of two classically correlated photons at 810\,nm (dashed line).  }
\label{fig:PSF}
\end{figure}

 For a more objective comparison, cross-sections of a triple-slit object of 70$\mu$m line width imaged at low NA  are shown  in Fig.~\ref{fig:TripleSlit}. OCM imaging resolution surpasses coherent and incoherent light at 810\,nm and shows practically identical resolution as 405\,nm.
Imaging  a point of 25\,$\mu$m Gaussian waist radius realized by focussing the pump laser and classical light sources in the object plane $\Sigma'_o$, the PSF of the single lens imaging at low NA is compared in Fig.~\ref{fig:PSF}. The theoretical curves for classical imaging are given by Eq.~\eqref{Eq:ImgCoh} and the classical PSF of Eq.~\eqref{Eq:SingleLensH} with $N=1$. Classically correlated photons pairs of OCM type would produce a centroid PSF  $|H(\bm X)|^2 = (|h|^2\ast|h|^2)(2\bm X)$ derivable analogously to Eq.~\eqref{Eq:Hrho}, see supplemental material~\cite{SuppMat}. This distribution is shown  for comparison (dashed line) and scales at the SQL. This limit is clearly beaten by the OCM PSF and thereby verifies the quantum correlation nature of the enhancement.

\section{Discussion}

The mechanism giving rise to the super-resolution can be understood by considering the far-field correlations of the OCM state. As evident in Eq.~\eqref{Eq:PsiFF}, it shows strong wavevector correlations or, equivalently, position correlations in the pupil plane. This situation can be regarded as a multi-mode NOON-state emanating from the lens and propagating to the image plane. As shown in \cite{Shin11, Rozema14} for two-mode NOON states, this leads to  $N$-times narrower multi-photon interference fringes in the image plane.  
The here presented multi-mode case allows to build up an image by the coherent superposition of such fringes of different directions and sizes.

Instead of using the proposed experimental setting to generate the OCM state, other means are possible. To avoid the state generation part of the setup, a thin NLC can be placed directly at  the imaging lens covering the full pupil area. For improvement of photon number $N$, higher order non-linear effects could be used \cite{Corona11}. Furthermore, the following theoretically proposed methods of resolution enhancement, at first glance unrelated to OCM, show the same fundamental mechanism.

As previously mentioned, super-resolution can be achieved by using a spatially correlated light source for illuminating the object and using a $N$-photon transmitter (NPT) in front of the imaging lens  \cite{Gio09}. This device transmits spatially correlated $N$-photon states only, but maintains the $N$-photon coherence. Effectively, this filtering projects the incoming state to Eq.~\eqref{Eq:PsiFF} or its incoherent imaging analogue and would therefore create an OCM state in the image plane. In the presented experiment, the NPT in front of the imaging lens would fully transmit the prepared OCM state and is therefore unnecessary. The general principle can be extended to an optical system of many lenses, where a NPT is placed at every lens. Intuitively, this mimics the propagation of a single photon of the de~Broglie wavelength by forcing all $N$ photons to stay together and tracing the same paths.

One second approach was proposed in \cite{Shih07} for $N=2$. The object is illuminated by spatially correlated biphoton light. Ultra-fast temporal quantum correlations of the pair is used trough their relative time-of-arrival in the image plane to post-select on correlated positions on the lens pupil, where both photon take the same path. In addition to an improvement of different origin, resolution at the Heisenberg limit is derived. This procedure effectively post-selects the state in Eq.~\eqref{Eq:PsiFF} or its incoherent imaging analogue. It still has to be analysed how temporal and spatial resolution are related.

\medskip

In conclusion, our theoretical and experimental results demonstrate that quantum states of light  showing super-resolution at the Heisenberg limit can be engineered. By limiting the Rayleigh resolution in low NA single-lens imaging, different light sources are compared in their ability to transmit spatial information. The OCM biphoton  state used in our experiment shows a resolution enhancement close to a factor of two and is  comparable to imaging at half the wavelength. For high NA systems, where the classical resolution is mainly limited by the wavelength, or for higher photon number $N$, theory suggests the possibility to have sub-wavelength image features present in the centroid coordinate. A full vectorial field analysis in contrast to the scalar approximations has yet to  show the advantage in the limit of high NA. 

Integrated single-photon detector arrays as presented here will certainly give rise to more experiments and applications in the field of quantum imaging. While the device in this work has non-optimal detection efficiency at the used wavelength, a  speed up in acquisition time and higher photon number correlation measurement is expected in more optimized settings. 

As elaborated by Tsang in \cite{Tsang09} and shown here, the image acquisition efficiency of the OCM correlations is very high due to the fact that every $N$-photon event carries image information. In the case of an $N$-photon absorbing film in lithography, where only spatially coincident events ($\bm\xi_k=0,\;k \in \{1,\ldots,N\}$) are registered, the image would be reproduced at $N$-fold resolution but at an efficiency which is expected to drop exponentially with $N$ 
\cite{Gio09}.  Efficiency can be gained by loosening the condition of strong correlation in far-field. An analysis of resolution and efficiency versus correlation length and photon number has yet to be performed.

The here developed theory  is general and allows for different experimental realizations. As a concrete example, the presented biphoton experiment serves as a proof-of-concept, clearly agreeing to the theory and showing the advantage over classical imaging schemes. Moreover, by unifying the understanding of recent results aiming at super-resolution at the Heisenberg limit, new paths of research for progressing in quantum imaging are opened.


\section*{Acknowledgments}
We thankfully acknowledge the support of the European Commission through the SUPERTWIN project, id.~686731. This research was supported by the grant PP00P2\_159259 of the Swiss National Science Foundation.

%


\clearpage
\onecolumngrid
\appendix

\setcounter{equation}{7}
\setcounter{figure}{4}

\section{Supplemental Material}

\subsection{Derivation of the OCM Imaging Equation} \label{Sec:AppDerOCM}

This section provides a step-by-step derivation of Eq.~(3) and (4) of the main text. 
Starting from the $N$-th order correlation function in the image plane
\begin{equation*}
G^{(N)}(\bm\rho_1,...,\bm\rho_N) = \left| \int d^2\!\bm\rho_1' \ldots d^2\!\bm\rho_N' \; A\left(\frac{\bm\rho'_1+\ldots+\bm\rho'_N}{N} \right) h\left(\frac{\bm\rho_1}{m}-\bm\rho_1'\right) \ldots h\left(\frac{\bm\rho_N}{m}-\bm\rho_N'\right)  \right|^2,
\end{equation*}
where the simplification of setting the magnification to $m=1$  does not  limit the generality of the following result.
A coordinate change to the centroid and deviation variables
$$
	\bm X = \frac 1 N \sum_{k=1}^N \bm\rho_k, \quad \bm\xi_k = \bm\rho_k - \bm X, \; k\in\{1,\ldots, N\}
$$
yields for the integral in the modulus
\begin{equation*}
  N^2 \int d^2\bm X' d^2\bm\xi_1' \ldots d^2\bm\xi_{N-1}' \; A\left(\bm X' \right) h\left({\bm X+\bm\xi_1}-(\bm X'+\bm\xi_1')\right) \ldots h\left({\bm X+\bm\xi_N}-(\bm X'+\bm\xi_N')\right) .
\end{equation*}
The Jacobian determinant of this coordinate transformation can be readily shown to be $|\det J| = |\det J_x|\cdot |\det J_y| = N^2$ and is present as a prefactor. By a change of variables to $\bm\xi_k'' = \bm\xi_k - \bm\xi_k'$, we get
\begin{equation*}
   \int d^2\!\bm X' A\left(\bm X' \right) \left( N^2 \int d^2\bm\xi_1'' \ldots d^2\bm\xi_{N-1}'' \, h\left(\bm\xi_1'' + \bm X - \bm X'\right) \ldots h\left(\bm\xi_N'' + \bm X - \bm X'\right) \right).
\end{equation*}
The term in parentheses is a function $H(\bm X - \bm X')$ and serves as an effective OCM PSF. Therefore
$$H(\bm X) = N^2 \int d^2\bm\xi_1 h(\bm\xi_1+\bm X) \ldots \int d^2\bm\xi_{N-1}\, h(\bm\xi_{N-1}+\bm X) \,h\left(\bm\xi_N + \bm X \right).
$$
Due to $\sum_k \bm\xi_k = 0$ we can infer $\bm\xi_N = -\sum_{k=1}^{N-1} \bm\xi_k$. By  defining the notation $h^{\ast k} \dot{=} (h \ast \ldots \ast h)$ for the $k$-times repeated self-convolution of a function $h = h^{\ast 1}$, we can deduce
\begin{align*}
H(\bm X) &= N^2 \int d^2\bm\xi_1 \, h(\bm\xi_1+\bm X) \ldots \int d^2\bm\xi_{N-1}\, h(\bm\xi_{N-1}+\bm X) \,h\left(\bm X - \sum_{k=1}^{N-1} \bm\xi_k \right)\\
&= N^2 \int d^2\bm\xi_1 \,h(\bm\xi_1+\bm X) \ldots 
\underbrace{\int d^2\bm\xi_{N-1}' \, h(\bm\xi_{N-1}') \,h\left(2\bm X - \sum_{k=1}^{N-2} \bm\xi_k - \bm\xi_{N-1}' \right)}_{h^{\ast 2} \left(2\bm X - \sum_{k=1}^{N-2} \bm\xi_k \right)} \\
&= N^2 \int d^2\bm\xi_1 \,h(\bm\xi_1+\bm X) \ldots   \int d^2\bm\xi_{N-2} \, h(\bm\xi_{N-2} + \bm X) \,h^{\ast 2}\left(2\bm X - \sum_{k=1}^{N-2} \bm\xi_k \right)\\ 
&=\ldots 
= N^2 \int d^2\bm\xi_1 \, h(\bm\xi_1+\bm X) \,h^{\ast (N-1)}\left((N-1)\bm X - \bm\xi_1 \right)\\
&= N^2 \, h^{\ast N} (N\bm X)
\end{align*}
This is the result of Eq.~(4). The OCM PSF can be written as $N$-times repeated self-convolution of the  PSF $h(\bm\rho)$ of the optical system.
For a general magnification factor $m$, we have therefore derived 
\begin{align*}
G^{(N)}(\bm X, \bm\xi_1, \ldots, \bm\xi_{N-1}) = \left| \int d^2\!\bm X'  \; A\left(\bm X' \right) H\left(\frac{\bm X}{m}-\bm X'\right) \right|^2 = \left|(A \ast H)\left(\frac{\bm X}{m}\right)\right|^2  .
\end{align*}

\subsection{Incoherent OCM Imaging}\label{App:IncohOCM}

The OCM state of Eq.~(2) with a sharp centroid position at location $\bm X_0$ can be gained by replacing $A(\bm \rho) \rightarrow A_{\bm X_0}(\bm X) = \delta^{(2)}(\bm X-\bm X_0)$. This state shall be denoted by 
$$\ket{\Psi_{\bm X_0}} = \int d^2\!\bm\rho_1 \ldots d^2\!\bm\rho_N \; \delta^{(2)}\left(\frac{\bm\rho_1+\ldots+\bm\rho_N}{N} - \bm X_0\right)\ket{\bm\rho_1, \ldots,  \bm\rho_N}.$$
For an object aperture function $A(\bm\rho)$, the mixed state given by the density operator
$$ \rho = \int d^2 \! \bm X \, |A(\bm X)|^2 \,  \ket{\Psi_{\bm X}}\bra{\Psi_{\bm X}} $$
contains an incoherent image. It is straightforward to show using the results of section~\ref{Sec:AppDerOCM} that with this mixed state as an input in plane $\Sigma_o$ of  setup in Fig.~1, the correlation function in $\Sigma_i$ will read
$$G^{(N)}(\bm X, \bm\xi_1, \ldots, \bm\xi_{N-1}) =  \int d^2\!\bm X' \left| \; A\left(\bm X' \right) H\left(\frac{\bm X}{m}-\bm X'\right) \right|^2 = (|A|^2 \ast |H|^2)\left(\frac{\bm X}{m}\right).
$$
This is formally analogues to classical incoherent imaging. The image is formed point-by-point, no interferences can occur. 

This imaging could be realised in our experimental implementation by focusing the pump beam in the object plane $\Sigma_o'$ and randomly scanning over the aperture. This classical randomness would produce the mixed state described above. Tsang proposed in \cite{Tsang09} such a scheme using a quantum ``laser pointer'' to build up an image incoherently at super-resolution. A disadvantage is that the single-shot property of the state generation would be lost in such an approach.

\subsection{OCM State Generation using SPDC }

This section derives Eq.~(7), the biphoton state at the output plane of the state preparation. 
Assuming narrow-band pumping and fixed detection wavelengths realized by spectral filtering, the generated biphoton state at the  central plane of the NLC reads
\begin{equation}\label{Eq:AppPsiSPDC}
	\ket{\Psi} = \int d^2\bm q_s d^2\bm q_i \, \tilde{E}_p\left(\bm q_s+ \bm q_i\right)\,
	\operatorname{sinc}\left(\frac{  \Delta k \,L}{2} \right) \ket{\bm q_s,\omega_s}\ket{\bm q_i, \omega_i}
\end{equation}
in transverse wavevector coordinates $\bm q_s$ and $\bm q_i$ for the signal and idler photon at their corresponding angular frequencies $\omega_s$ and $\omega_i$, and the pump field distribution $\tilde E_p(\bm q)$ at angular frequency $\omega_p$ \cite{Monken98}. Energy conservation imposes $\omega_p = \omega_s + \omega_i$.
The wavevector mismatch 
$$ \Delta k = \sqrt{\vphantom{\left(\frac{\omega_s+\omega_i}{c}n(\omega_s+\omega_i)\right)^2}
 \left(\frac{\omega_s}{c}n(\omega_s)\right)^2 - \bm q_s^2} 
 + \sqrt{ \vphantom{\left(\frac{\omega_s+\omega_i}{c}n(\omega_s+\omega_i)\right)^2}
 \left(\frac{\omega_i}{c}n(\omega_i)\right)^2 - \bm q_i^2} 
 - \sqrt{\left(\frac{\omega_s+\omega_i}{c}n(\omega_s+\omega_i)\right)^2  - (\bm q_s+ \bm q_i)^2} + \frac{2\pi}{G}$$
where $c$ is the speed of light and the refractive index $n(\omega)$ of the crystal is given by its temperature dependant Sellmeier equations.  The NLC poling period $G$ is fixed at its fabrication and is chosen to achieve $\Delta k = 0$ at the used wavelengths and for collinear emission $\bm q=0$. 

Let the function $E_o(\bm\rho)$ define the electric field distribution of the monochromatic pump of angular frequency $\omega_p$ in the object plane $\Sigma_o'$ of Fig.~1. Propagating it trough the first lens of focal length $f$ to the center of the crystal, the far-field plane relative to $\Sigma_o'$, the field distribution is given  by $ E_p(\bm \rho) = \tilde E_o \left( {\omega_p\bm \rho}/{c f} \right)$ with the Fourier transform $\tilde E_o (\bm q) = \int d\bm\rho \, E_o(\bm\rho)\, e^{-i\bm q \bm \rho}$
 \cite{Good88}. Therefore, this pump field incident on the NLC has a Fourier transform of
\begin{equation}\label{Eq:AppEp}
 \tilde E_p(\bm q)  = E_o\left(-\frac{c f}{\omega_p} \bm q \right)
\end{equation}
and can be inserted in Eq.~\eqref{Eq:AppPsiSPDC}.

The SPDC state has to be propagated from the NLC through the lens of focal length $f$ to the output plane $\Sigma_o$. The latter is the far-field plane relative to the NLC central plane. Plane waves of transversal wavevector $\bm q$ emitted by the NLC are focused to a location $\bm\rho = \frac{c f}{\omega}\bm q$  \cite{Good88}. Formally, the state propagation can be performed by the transformation $\ket{\bm q,\omega}\rightarrow\ket{\frac{cf}{\omega}\bm q=\bm\rho,\omega}$. Using Eq.~\eqref{Eq:AppEp} and Eq.~\eqref{Eq:AppPsiSPDC}, we get 
\begin{equation}\label{Eq:AppPsiNF}
	\ket{\Psi} = \int d^2\bm \rho_s d^2\bm\rho_i \, 
	{E}_o\left(-\frac{\omega_s\bm\rho_s + \omega_i\bm\rho_i}{\omega_s+\omega_i}  \right)\,
	\operatorname{sinc}\left(\frac{  \Delta k \,L}{2} \right) \ket{\bm \rho_s,\omega_s}\ket{\bm \rho_i, \omega_i}
\end{equation}
where $\Delta k$ is evaluated at $\bm q_k = (\omega_k/cf)\bm\rho_k, \; k \in \{s,i\}$. For frequency degenerated emission with $\omega_s=\omega_i=\tfrac 1 2\omega_p$,  the biphoton state in the preparation output plane $\Sigma_o$ reads
\begin{equation*}
	\ket{\Psi} = \int d^2\bm \rho_s d^2\bm\rho_i \, 
	{E}_o\left(-\frac{\bm\rho_s + \bm\rho_i}{2}  \right)\,
	\operatorname{sinc}\left(\frac{  \Delta k \,L}{2} \right) \ket{\bm \rho_s,\tfrac 1 2 \omega_p}\ket{\bm \rho_i,\tfrac 1 2 \omega_p}.
\end{equation*}
Finally, uniformly illuminating an object aperture $A(\bm\rho)$ in the plane $\Sigma_o'$ yields $E_o(\bm\rho) = A(\bm \rho)$.

Eq.~\eqref{Eq:AppPsiNF} shows the necessity to filter for frequency degenerate emission in order to properly reconstruct the image with the centroid. Broadband emission could be conceived if the detection provides spectral information.

\subsection{Standard Quantum Limit of Classically Correlated Photons}\label{App:ClOCM}

This section derives a centroid PSF for classically correlated photons with no entanglement and shows its behaviour at large photon number $N$. For simplicity, we set first $N=2$. Assuming OCM-like correlations for a point object, in the object plane these are modelled by a classical, multi-variate probability density  function
$$ p_o(\bm\rho_1, \bm\rho_2) = \delta^{(2)}(\bm\rho_1+\bm\rho_2).$$
Propagating the photons through the resolution limited imaging system, the process of blurring by the PSF can be described by adding a random variable $\bm N $ to their original positions in the object plane
$$ \bm\rho_k \rightarrow \bm\rho_k + \bm N_k, \quad k \in \{1,2\}. $$
As both photons are affected by the PSF independently, a separate random spread term  is needed for each. Their probability density function is given by the PSF of the imaging system with $P_{\bm N_k}(\bm N) = |h(\bm N)|^2$.
Using elementary statistics, we get for the probability density in the image plane 
\begin{align*}
 p_i(\bm\rho_1, \bm\rho_2) &= \int d^2\!\bm\rho_1' d^2\!\bm\rho_2' \, p_o(\bm\rho_1,\bm\rho_2) \, |h(\bm\rho_1-\bm\rho_1')|^2 \, |h(\bm\rho_2-\bm\rho_2')|^2 \\
& = \left( |h|^2 \ast |h|^2\right) (\bm\rho_1 + \bm\rho_2).
\end{align*}
This is formally equivalent to independent, incoherent imaging of the photons and therefore validates the statistical model assumed. Using $\bm\rho_- = \tfrac 1 2 (\bm\rho_1 -  \bm\rho_2)$, the centroid random variable $\bm X_i$ in the image plane has a probability density function
$$ P_{\bm X_i}(\bm X) = \int d^2\!\bm\rho_- \, p_i(\bm X + \bm\rho_-, \bm X - \bm\rho_-) = \int d^2\!\bm\rho_- \, \left( |h|^2 \ast |h|^2\right) (2\bm X) \propto \left( |h|^2 \ast |h|^2\right) (2\bm X).$$

Generalizing to a classically correlated $N$-photon state, on can use the fact that the centroid random variable in the image plane is 
$$ \bm X_i = \frac 1 N \left(\sum_{k=1}^N \bm\rho_k + \bm N_k \right).$$
With the OCM property $\sum_{k=1}^N \bm \rho_k = 0$ of a point object at the origin, we can conclude that the probability distribution of $\bm X_i$  is given by the sum of $N$ independent noise sources and reads
\begin{equation}\label{Eq:AppClOCM}
 P_{\bm X_i}(\bm X) = \left( |h|^2 \right)^{\ast N}(N\bm X) 
\end{equation}
using the self-convolution notation from Sec.~\ref{Sec:AppDerOCM}. This is the PSF of classical OCM imaging and the optimum for any imaging using classically correlated photons.

Because the noise terms are independent, identically distributed random variables, we can apply the central limit theorem of probability theory. This says, that the probability density of Eq.~\eqref{Eq:AppClOCM} converges in the limit of large $N$ to a normal distribution 
$$ P_{\bm X_i}(\bm X) \xrightarrow{N\rightarrow \infty}  \frac 1 {\sqrt{2\pi\sigma^2/N}} \,e^{-\frac{\bm X^2}{2\sigma^2/N}}$$
with the standard deviation  $\sigma$ of  the  imaging system PSF  $|h(\bm\rho)|^2$. The asymptotic $1/\sqrt{N}$ decrease in width indicates a resolution enhancement at the SQL.

\subsection{Near- and Far-Field Correlation Measurements}

\begin{figure}[tp]
	\centering
	\subfigure[Near-Field Correlations]{\label{fig:NF_XCorr}
		\includegraphics[height=3.9cm]{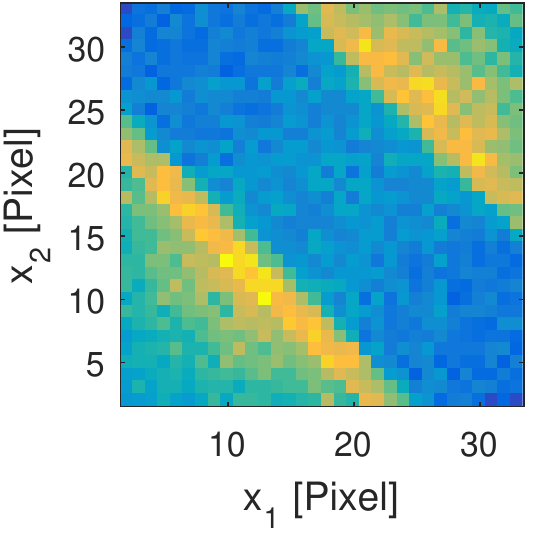}}\hspace{.1cm}
	\subfigure[Far-Field Correlations ]{\label{fig:FF_XCorr}
		\includegraphics[height=3.9cm]{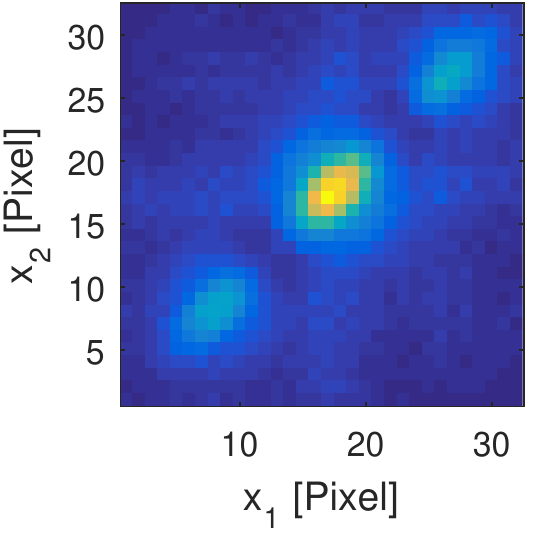}
		\raisebox{.8cm}{\includegraphics[trim= 0 0 0 0, height=3.2cm]{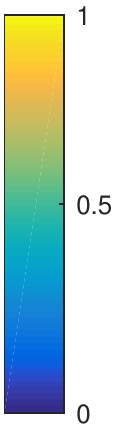}}} \hspace{.1cm}
	\subfigure[Far-Field, OCM Diagonal vs. 405\,nm]{\label{fig:FF_XCorr_diag}
		\includegraphics[height=3.9cm]{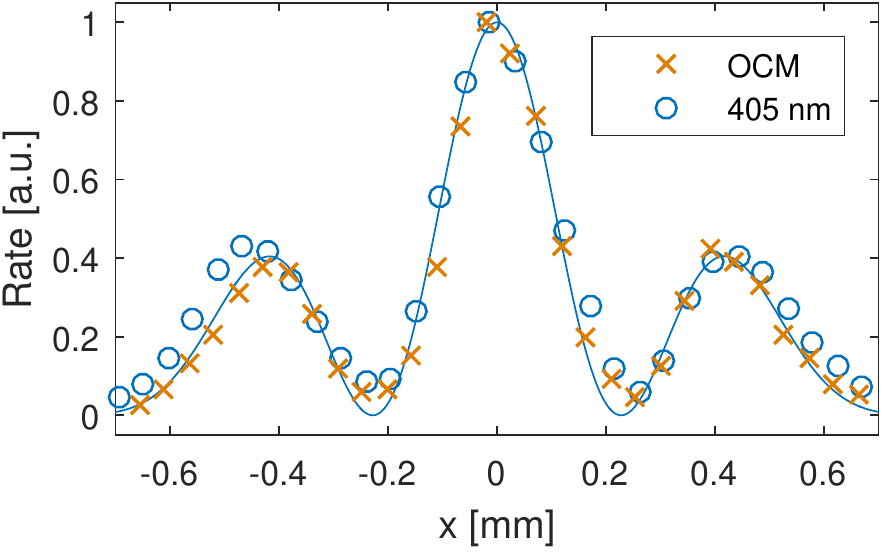}}   
	\caption{The biphoton OCM state of a double-slit is analyzed in near- and far-field for position-correlations  in $x$-direction orthogonal to the lines. The high NA near-field measurement in~(a) shows the image features on the diagonals and thereby in the centroid position. In far-field~(b), strong correlation is observed. (c) shows the diagonal of this OCM diffraction pattern (orange crosses) and  the expected narrowing  down to the width produced with coherent light at 405\,nm (blue circles). The theoretical curve for the latter is shown. }
	\label{fig:NF_FF}
\end{figure}

With the full measurement of the correlation function in Eq.~(3), spatial correlations of the OCM state can be analyzed. A double-slit of 200\,$\mu$m line width is used as object. The OCM state shows in Fig.~\ref{fig:NF_XCorr} position correlations along the $x$-direction orthogonal to the slits. This can be understood in terms of a \mbox{1-D} version of  Eq.~(3) with $N=2$, where the image is encoded in the centroid $\bm X$ (diagonal) and independence in $\bm\xi_1$ (anti-diagonal) is present. Both inner slit edges and the central dark separation are visible, vignetted by the pump beam shape. Replacing the imaging lens $L_3$ in the setup by a far-field lens of 400\,mm focal length  placed in this distance in front of the detector, far-field correlations of Fig.~\ref{fig:FF_XCorr} are measured. Strong position correlations are observed in agreement with Eq.~(6). Furthermore, comparing the diagonal of this biphoton diffraction pattern to a measurement with coherent light at 405\,nm in Fig.~\ref{fig:FF_XCorr_diag} yields almost identical results, confirming the theoretically predicted relevance of the twice smaller de~Broglie wavelength. A related result was obtained in \cite{DAngelo01}.

\end{document}